# Both electrical and metabolic coupling shape the collective multimodal activity and functional connectivity patterns in beta cell collectives: A computational model perspective

Marko Šterk[1,2,3], Uroš Barać[1], Andraž Stožer[2], Marko Gosak[1,2,3,*]

[1] Department of Physics, Faculty of Natural Sciences and Mathematics, Koroška cesta 160, University of Maribor, 2000 Maribor, Slovenia

[2] Institute of Physiology, Faculty of Medicine, Taborska ulica 8, University of Maribor, 2000 Maribor, Slovenia

[3] Alma Mater Europaea, Slovenska ulica 17, 2000 Maribor, Slovenia

* marko.gosak@um.si

Pancreatic beta cells are coupled excitable oscillators that synchronize their activity via different communication pathways. Their oscillatory activity manifests itself on multiple timescales and consists of bursting electrical activity, subsequent oscillations in the intracellular $Ca^{2+}$, as well as oscillations in metabolism and exocytosis. The coordination of the intricate activity on the multicellular level plays a key role in the regulation of physiological pulsatile insulin secretion and is incompletely understood. In this contribution, we investigate theoretically the principles that give rise to the synchronized activity of beta cell populations by building up a phenomenological multicellular model that incorporates the basic features of beta cell dynamics. Specifically, the model is composed of coupled slow and fast oscillatory units that reflect metabolic processes and electrical activity, respectively. Using a realistic description of the intercellular interactions, we study how the combination of electrical and metabolic coupling generates collective rhythmicity and shapes functional beta cell networks. It turns out that while electrical coupling solely can synchronize the responses, the addition of metabolic interactions further enhances coordination, the spatial range of interactions, increases the number of connections in the functional beta cell networks, and ensures a better consistency with experimental findings. Moreover, our computational results provide additional insights into the relationship between beta cell heterogeneity, their activity profiles, and functional connectivity, supplementing thereby recent experimental results on endocrine networks.

# I. INTRODUCTION

Pancreatic beta cells within the pancreatic islets of Langerhans possess an inherent ability to generate multimodal oscillatory signals as a response to changes in the extracellular concentrations of glucose and other nutrients, as well as hormones and neurotransmitters [1]. This oscillatory activity is characterized by tightly coupled membrane potential and intracellular $Ca^{2+}$ concentration dynamics that are intertwined with oscillations in metabolism [2–5] and drive the pulsatile exocytosis of insulin [6,7]. The rhythmic pattern of insulin delivery to the target tissues is essential for ensuring their sensitivity and metabolic homeostasis, whilst its disruptions are associated with diabetes mellitus [8,9]. The oscillatory activity of beta cells manifests itself on different timescales and results from networked feedback interactions of various oscillatory subsystems, such as the glycolytic, mitochondrial, and electrical/calcium components [10]. The slow oscillations, likely indicative of metabolic activity, have a frequency of 0.06–0.2 $min^{-1}$, aligning with the temporal scale of plasma insulin oscillations [11–14]. Superimposed on these slow oscillations are the so-called fast $Ca^{2+}$ oscillations with a frequency of about 5 $min^{-1}$ and a duration of about 2–15 s, which reflect the bursting pattern of electrical activity. These oscillations are glucose-dependent [15–19] and govern presumably the amplitude of the slow plasma insulin oscillations, whereas their period does not seem to be glucose-dependent [8,20,21].

Ensuring proper insulin release patterns is a collective effort of beta cell collectives, which exist as morphologically well-defined socio-cellular units. It has been shown that coordinated multicellular activity is required for normal pulsatile secretion [12,22] and that intercellular communication can potentiate secretory output [23,24]. Moreover, intercellular coordination was not only found instrumental for appropriate insulin release, but also implicated in altered islet function, disturbances in metabolic homeostasis and plasma insulin oscillations, similar to what is observed in metabolic diseases [8,25,26]. For that reason, intra-islet signaling mechanisms are increasingly recognized as one of the underlying factors which contribute to the pathogenesis of diabetes [26–28]. While beta cells communicate among themselves and with other endocrine cells by different means [29,30], the coupling provided by gap junctions represents the main synchronizing mechanism. Gap junctions are a type of specialized membrane channels composed of connexin 36 proteins and allow communication through current-carrying ions and other small molecules transported between the cytoplasms of adjacent cells [31]. Ionic flow through gap junctions is the main determinant of cell-to-cell communication during the active phases and represents the key mediator of intercellular waves [32,33]. The flow of other larger molecules, such as nucleotides and intermediate products of glycolysis, might impact on pre-threshold intercellular communication and is presumably responsible for the synchronization of the slow oscillatory component [21,34,35,60]. However, which molecular species diffuses across gap junctions, how exactly metabolic coupling contributes to the alignment of the slow oscillations, and whether it plays a role in amplifying insulin secretion is not entirely clear and is a matter of current research.

Besides functioning as multimodal oscillators that interact in a non-trivial way, the beta cells are also highly heterogeneous [36,37]. Their heterogeneity manifests itself on different levels and has important functional implications, including the presence of specialized subpopulations that govern the responses to a changing environment [16,38–41]. To assess and quantify the principles that govern the collective activity in islets, advanced experimental methods, such as multicellular imaging, have recently been complemented with network analyses [24]. Describing interactions within beta cell collectives with network language has revealed that the underlying functional networks have a modular structure and are much more heterogeneous than one would expect from a gap junction-coupled syncytium [16,42,43]. The identified fingerprints of small-worldness and a broad-scale character imply the existence of highly connected cells, called hubs. While their exact role is still not entirely clear, it has been argued that they have some unique characteristics, which endows them with a higher-than-average influence on the multicellular activity within the islets [41,44–46]. Moreover, the hub cells also may be preferentially vulnerable to diabetogenic conditions, suggesting a possible mechanism behind the loss of pulsatile insulin secretion in diabetes [14,43,47]. Furthermore, the complex dynamical nature of beta cell collectives and the presence of specialized subpopulations was recognized as theoretically very appealing and has stimulated the design of multicellular beta cell models. The use of mathematical models has already provided additional insight into the mechanisms behind beta cell synchronization [2,48–50], propagation of intercellular waves [32,51], and on how coherent collective activity emerges from heterogeneous subpopulations [52–56]. Recently, computational models have also been used to investigate the influence of different physiological determinants on the characteristics of functional beta cell networks [41,45,57–60]. *In silico* studies have also been utilized to assess how different types of intercellular coupling contribute to synchronized behavior within the islets. Specifically, it has been shown that diffusion of metabolic intermediates synchronizes the slow oscillatory activity and enhances the coordinated behavior of the fast activity, although the metabolic coupling was found not to be strictly necessary for the alignment of electrical activity [13,35,61]. Interestingly, Loppini and Chiodo have shown that metabolic coupling does not only foster synchronous activity of the slow oscillatory component, but also that the extent of synchronization is further enlarged by the presence of hub cells [62]. Nevertheless, to what extent different means of intercellular communication shape the complex spatiotemporal activity in islets, is incompletely understood and remains to be clarified.

In the present study, we investigate numerically how the combined influence of electrical and metabolic coupling gives rise to coordinated activity patterns in beta cell populations. To this purpose, we designed a phenomenological beta cell model composed of coupled slow and fast oscillatory units. In contrast to more exhaustive computational models, the small number of parameters included in our minimalistic framework eases a systematic and straightforward analysis. To model intercellular interactions, we implement a realistic coupling scheme that is based on experimental data. In our investigation we focus on how the combination of electrical and metabolic coupling generates collective multimodal rhythmicity and influences the functional beta cell networks. We also evaluate the similarities between the structural and functional networks and assess our findings in the light of the latest experimental discoveries.

## II. MATHEMATICAL MODEL AND ANALYSIS

### A. Single cell model

We use a phenomenological multicellular model to simulate the multimodal nature of oscillatory activity in a network of beta cells. Specifically, we propose a double-layered minimal model to describe the interplay between the fast electrical activity and the slow metabolic processes. By this means, the dynamics of each beta cell, i.e., each node in the network, is governed by two oscillators: the Poincaré oscillator to model the slow oscillatory component and the two-dimensional iterated Rulkov map to simulate the fast electrical activity. The differential equations for the $i$-th Poincaré oscillator in the Cartesian coordinates are as follows:

$$\dot{x}_i = -y_i\omega_i - \gamma x_i\left(\sqrt{x_i{}^2 + y_i{}^2} - A\right) + K_P \sum_{j \in S_i} w_{ij}(x_j - x_i), \quad (1)$$

$$\dot{y}_i = x_i\omega_i - \gamma y_i\left(\sqrt{x_i{}^2 + y_i{}^2} - A\right) + K_P \sum_{j \in S_i} w_{ij}(y_j - y_i), \quad (2)$$

where $\gamma = 1.0$ is the relaxation rate, $A = 0.2$ is the amplitude, $K_P$ the coupling strength between Poincare oscillators, $w_{ij}$ the edge weight between node $i$ and $j$, $S_i$ indicates all neighbors of node $i$, and parameter $\omega_i$ denotes the intrinsic frequency of the $i$-th oscillator. To account for the differences in metabolic activity observed in beta cells [37], values of $\omega$ were distributed among the oscillators in the range $\omega_{mean} \pm 0.15 w_{\text{avg}}$, so that the frequencies of the slow oscillatory component vary in principle $\pm 15\%$. Notably, we correlated the deviation for the average frequency with the sum of edge weights, as explained in more detail in continuation. The fast electrical activity is described by the Rulkov oscillator, which is written as an iterative map [63]:

$$u_i(n+1) = \frac{\alpha_i(n)}{1+u_i(n)^2} + v_i(n) + D\xi_i(n) + K_R \sum_{j \in S_i} w_{ij}(u_j(n) - u_i(n)), \quad (3)$$

$$v_i(n+1) = v_i(n) - \sigma_i u_i(n) - \chi_i, \quad (4)$$

where $n$ is a discrete time step and $u_i$ and $v_i$ are dimensionless variables that resemble the membrane potential and the gating variable, respectively. $K_R$ the coupling strength between Rulkov oscillators and $D = 0.005$ indicates the strength of Gaussian noise $\xi$ with mean 0 and variance 1 that accounts for stochasticity in beta cell activity. Parameters $\sigma$ and $\chi$ were randomly distributed to account for beta cell heterogeneity on the electrophysiological level, so that each $\sigma_i$ was selected randomly from the interval [0.001, 0.0014] with $\chi_i = \sigma_i$.

The parameter $\alpha_i$ was used to interconnect both oscillators: the $x_i$ amplitude of $i$-th Poincare oscillator drives the value of $\alpha_i$ and thereby the level of excitability of $i$-th Rulkov oscillator. Specifically, parameter $\alpha_i$ that characterizes the cellular excitability level [53,64] is defined as $\alpha_i = (x_i + A)(\alpha_{MAX} - \alpha_{MIN})/2A + \alpha_{MIN}$ so that the values of $\alpha_i$ oscillate in phase with $x_i$ in the range $\alpha_{MIN}$ to $\alpha_{MAX}$. The range of $\alpha$ is set from 1.95 to 1.995, whereby higher values $\alpha$ indicate a higher level of excitability and frequency of oscillations [53]. In this manner, we introduced a simplified description of the symbiotic relationship between the slow metabolic (Poincaré) and the fast electrical (Rulkov) oscillations, focusing solely on the influence of the metabolic oscillator on the electrical oscillator, although in reality this interaction is more intricate, involving reciprocal effects [10,65]. The composite cellular signal $c_i$ is derived by combining the contributions of both oscillators: $c_i = x_i + bu_i$, where factor $b = 0.5$ serves as a scaling parameter for the Rulkov component, $u_i$. This calculation results in a simulated signal that closely resembles experimentally measured signals, as exemplified in Fig. 1(a), where we present a representative simulated trace of beta cell activity, which is generated from the dynamic interplay of both oscillators. Note that that the composite signal is used solely for visualization purposes, while for subsequent analyses, both dynamical components were examined individually.

Although the Rulkov iterative model was originally proposed as a phenomenological model of neuronal activity, its use is not limited to neuronal dynamics and can be applied to other excitable cells as well, including pancreatic beta cells [53]. Previously, we conducted an analysis of the Rulkov model and demonstrated that it resides in the excitability regime near alpha equals 2. In this regime, oscillations can occur in the variable $u$, reflecting the membrane potential value, with the oscillation frequency dependent on the alpha parameter's value [53]. Variations in parameter alpha thus reflect variations in the level of excitability, mimicking thereby changes in the conductance of glucose-sensitive $K_{ATP}$ channels, which serve as the primary triggers for beta cell behavior in detailed cellular models [66,67]. Furthermore, the electrical activity in beta cells is intricately intertwined with metabolic processes [10,65]. In our minimal model, we have introduced metabolic oscillations as an autonomous oscillator that influences the level of excitability. In this vein we use the Poincaré oscillator to replicate the series of intracellular metabolic processes, including the depletion of glycolytic intermediates in mitochondria and the subsequent oscillations in the ATP/ADP ratio, which in turn determine $K_{ATP}$ channel activity. Although our minimalist modeling approach doesn't provide mechanistic insights into physiological processes and signaling pathways, it offers the advantage of simplicity with only a few intuitive parameters. This makes the model well-suited for efficiently exploring the fundamental principles governing synchronous behavior in beta cells, particularly in the context of heterogeneity, intercellular communication and how it shapes the functional networks.

The Poincaré oscillator was numerically solved with Runge-Kutta second order and a timestep $dt = 0.05$ and since the Rulkov model is an iterative map we set $dt$ to equal one discrete time step $n$ in the Rulkov model. The $\omega_{\text{avg}}$ was then set to 0.006 so that the time of one cycle of the Poincaré oscillator corresponds to approximately 20 oscillations of the Rulkov oscillator and resembles the experimental observations [21]. For further analysis we integrated the system over approximately 50 cycles of the Poincare oscillator, whereby the initial 10 were discarded.

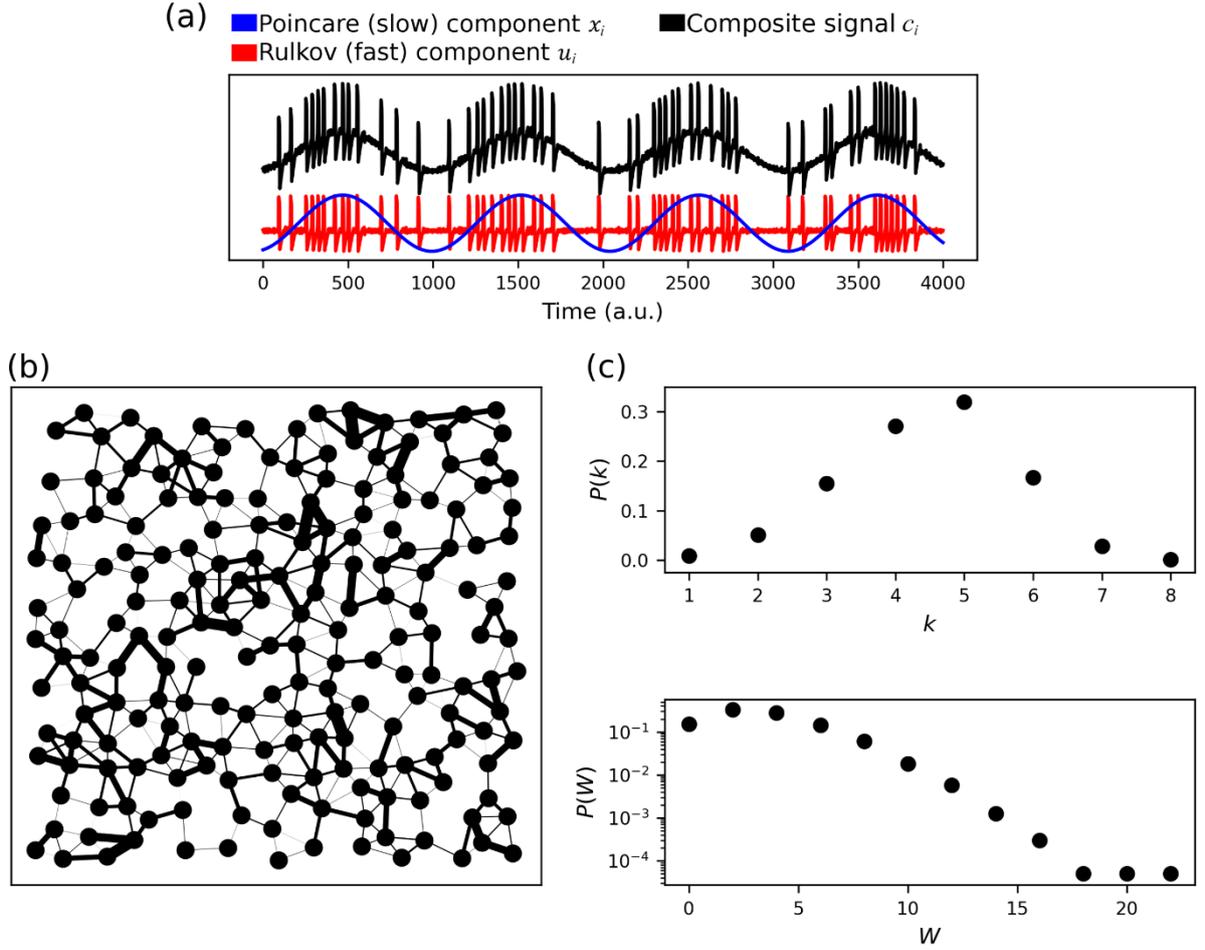

**FIG. 1. Characteristics and behavior of the phenomenological model of beta cell population.** (a) A simulated trace of beta cell activity (composite signal, black) that is composed of the slow component (Poincaré oscillator, blue) and the fast component (Rulkov map, red). (b) Network model of beta cell syncytium. Nodes represent individual beta cells and connections stand for gap junctional coupling. The widths of connections reflect the junctional conductances. (c) Degree distribution $P(k)$ (i.e., the distribution of number of connections per cell; upper panel) and the distribution of total weights of connections ($P(W)$; lower panel).

### B. Network of beta cells and intercellular communication

To model intercellular interactions between beta cells we constructed a random geometric network. $N = 200$ nodes were arranged randomly within a unit square with a condition of a minimum distance of $0.75\sqrt{N}$ between each pair of nodes, to ensure a realistic spatial distribution of cells. Pair of nodes $i$ and $j$ were considered connected if the distance between them was less than a threshold distance of $D_{th} = \sqrt{\langle k \rangle/(N\pi)}$, where $\langle k \rangle = 6$ is the average degree of the network. Additionally, each connection was weighted to incorporate heterogeneity in the intercellular coupling. The weight $w_{ij}$ [see Eqs. (1)-(3)] between the *i*-th and *j*-th oscillator were drawn randomly for an exponential distribution with mean 1:

$$w_{ij} = -\ln(rand(0,1)). \tag{5}$$

In Fig. 1(b) we show a typical beta cell network structure with the widths of connections reflecting their weights $w_{ij}$. In Figure 1C the degree distribution and the distribution of weights *w* are presented. The degree distribution (upper panel) solely quantifies the varying number of connections per cell, whilst the distribution of weighted node degrees (lower panel) additionally encompasses the coupling strengths, which reflects gap junctional conductances. It can be noticed that the number of neighbors spans between 1 to 8 with an average between 5 and 6, which is in accordance with experimental observations [68]. Moreover, the distribution of weights is rather heterogeneous, with an order of magnitude difference between the values, which also corresponds to variability in gap-junctional conductances observed in experiments [69]. Furthermore, in response to recent experimental data which shows that hub cells display higher metabolic rates [37,43], as well as a higher gap-junctional connectivity [46], we correlated the frequency distribution of the slow oscillatory component with the strength of gap-junctional coupling. Specifically, cells with the highest sum of neighboring edge weights were assigned the highest $\omega$ (i.e., $1.15\omega_{avg}$), whilst the cells with the lowest sum of neighboring edge weights were assigned the lowest $\omega$ (i.e., $0.85\omega_{avg}$). Parameters $\chi_i$ and $\sigma_i$ were assigned to cells at random, i.e., independent of the frequency of the slow component and the distribution of weights in the intercellular network model.

### C. Functional network construction and analysis

To assess the collective activity of cells we computed the Pearson correlation coefficient between the *i*-th and *j*-th cell, separately for the fast (Rulkov) and slow (Poincaré) oscillatory component as follows:

$$R_{i,j} = \frac{\sum_{n=1}^{N_p}(f(n)_i - \bar{f}_i)(f(n)_j - \bar{f}_j)}{\sigma_i \sigma_j}, \tag{6}$$

where *f(n)* represents the traces from either the fast, $u(n)$, or the slow, $x(n)$, oscillatory component at iteration step *n*, respectively, $N_p$ is the number of points in time series, and $\sigma$ denotes the corresponding standard deviations. By computing the Pearson correlation coefficient for all cell pairs in each simulation we constructed correlation matrices, which were then thresholded to construct functional networks, i.e., the *i*-th and *j*-th cell were considered as connected if $R_{ij} > R_{th}$, where $R_{th}$ denotes the threshold. This threshold was computed iteratively with a pre-set target $k_{avg}$. Specifically, the connectivity threshold was varied until the extracted functional networks had the same average network node degree as the underlying structural network ($k_{\text{avg}} \approx 6$) with a 5% tolerance. It should be noted that in our previous works we have already shown that within reasonable limits the results of the network analyses are qualitatively independent of the value used for thresholding or the specified average node degree [21,46].

We quantified the extracted functional intercellular networks by using conventional tools from complex network theory, as previously described [24]. Specifically, we computed the node degrees, the degree distributions, and the clustering coefficients [70]. To evaluate the similarities between different network types we made use of multilayer network formalism and calculated the so-called Jaccard similarity coefficient as:

$$J_{\alpha,\alpha'} = \frac{|A_\alpha \cap A_{\alpha'}|}{|A_\alpha \cup A'_\alpha|};$$

$$\alpha, \alpha' \in [\text{structural}, \text{fast}, \text{slow}], \tag{7}$$

where *A* represents the set of edges in the network *α* or *α*'. This yields a similarity coefficient between 0 and 1, where 0 means there is no overlap between the two networks and 1 means the networks are identical.

To assess the cellular activity, the computed traces were discretized. The fast Rulkov-driven component was binarized, so that the values of $u_i$ when denoting the depolarization phase were set to 1, and to 0 otherwise. The binarized signals were then used to visualize electrical activity patterns as raster plots and to compute the frequency of fast oscillations. The frequency of slow oscillations was determined simply based on the number of determined maxima in the Poincaré-driven component.

### III. RESULTS

We begin by presenting the dynamical responses of our multicellular phenomenological model of beta cell activity. In Fig. 2, we show the results for different sets of coupling parameters, including the average signal of the entire network (composite signal *c*, first row), the raster plot of the binarized fast electrical activity (Rulkov component *u*, second row), traces from three individual cells (composite signal $c_i$, third row), and outtakes of the binarized fast electrical activity (Rulkov component) along with the average composite cellular signal (last row). For Fig. 2(a), we set weak electrical intercellular coupling ($K_R$=0.004) and omitted metabolic coupling ($K_P$=0.0). As a result, the cells were weakly synchronized, as evidenced by the relatively low amplitude of the average signal and the lack of ordered patterns in the raster plot. Moving to Fig. 2(b), we increased the electrical intercellular coupling ($K_R$=0.01) while maintaining zero metabolic coupling. This led to a clear increase in the coherence of the spatiotemporal activity, though the lack of metabolic interactions prevented a slow oscillatory component from being contained in the average signal, because the metabolic activities of individual cells were out of phase. For the calculations presented in Fig. 2(c), we additionally included the coupling in the metabolic component. In this case, the slow oscillations were rather aligned and hence encoded in the average signals. Additionally, coupling in the slow oscillatory activity (Poincare component) appeared to modulate and coordinate the fast oscillations (Rulkov component), as evidenced by increased activity and the highest amplitude of the average signal around the maximal values, in accordance with experimental findings [21]. Notably, in our simulations synchronized activity among cells is observed in the form of intercellular waves. Supplementary videos S1-S3 showcase animations of multicellular activity with parameter values identical to those in Fig. 2. When the intercellular coupling is weak, the intercellular waves appear erratic and involve smaller clusters of cells. However, they become more extensive and regular with stronger intercellular coupling, resembling spatio-temporal activity patterns observed in experiments with sustained glucose stimulation [32,33,53]. In what follows, we will investigate the combined effect of electrical and metabolic coupling on the collective rhythmicity in beta cell networks in further detail.

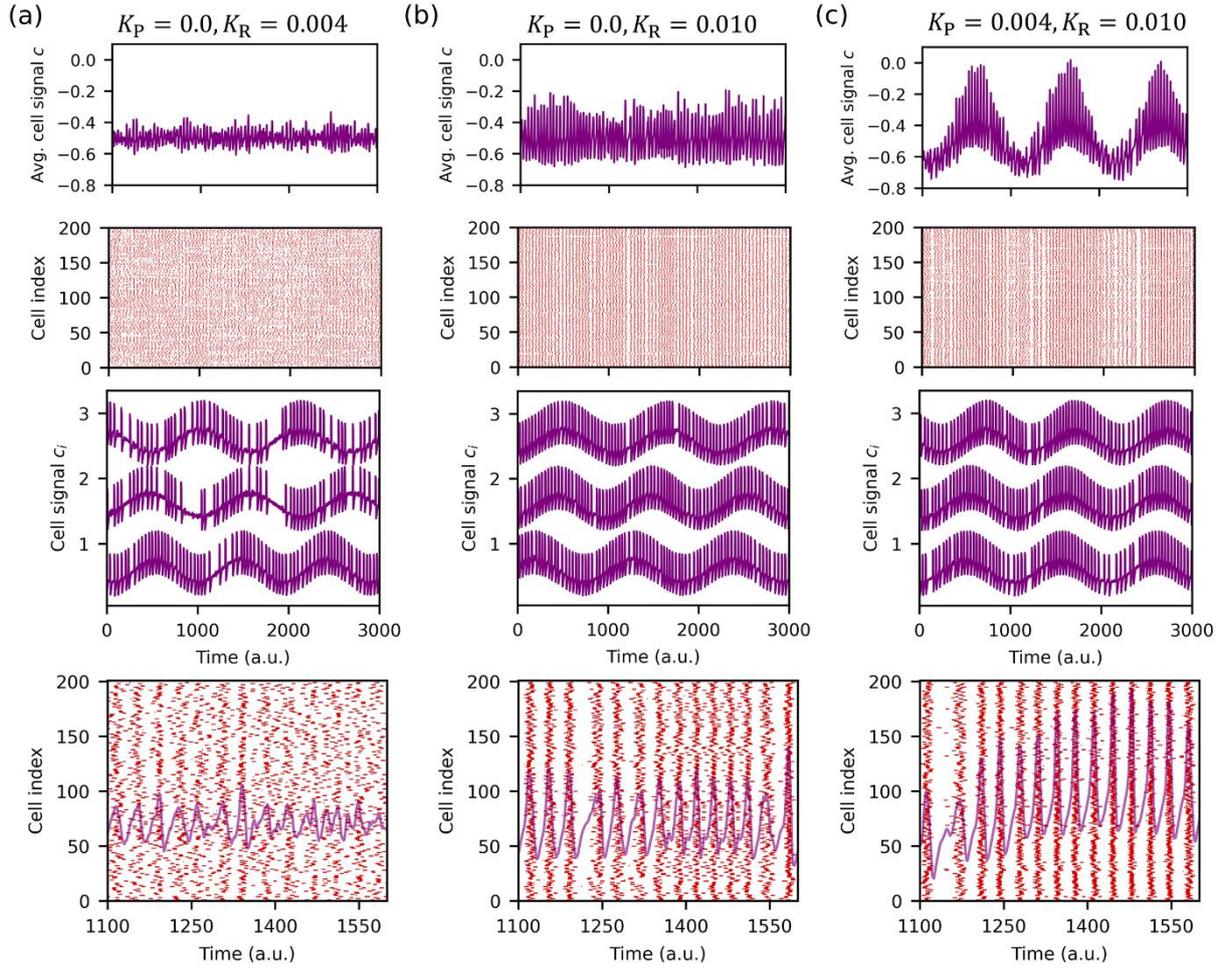

**FIG. 2. Intercellular coupling strength influences the coordinated beta cell activity.** Average signal of beta cell activity (composite signal $c$; first row), raster plot of binarized fast cellular activity (Rulkov component; second row), time series of three individual cells (composite signals $c_i$, third row), and outtake of the binarized fast cellular activity along with the average composite cellular signal (purple line on the top of red dots; last row) for different values of coupling strengths: $K_P=0.0$, $K_R=0.004$ **(a)**, $K_P=0.0$, $K_R=0.010$ **(b)**, and $K_P=0.004$, $K_R=0.010$ **(c)**.

We proceeded to investigate more explicitly how intercellular coupling affects the beta cell synchronicity and the resulting functional networks. In Fig. 3(a), we present color-coded values of the average correlation between fast electrical activity (Rulkov map) as a function of coupling strengths in the slow metabolic ($K_P$) and fast electrical ($K_R$) components. As expected, increasing $K_R$ leads to higher average correlations, but coupling in the slow oscillatory component also further enhances coherence in the fast component. Specifically, with intermediate coupling strength ($K_R=0.01$), which results in activity patterns like those observed in experiments, additional metabolic coupling can increase the average correlation between fast oscillations by up to 30%. To demonstrate the impact of metabolic coupling on the collective beta cell activity in further detail, we constructed functional connectivity networks based on simulated traces under different conditions. In Fig. 3(b), we show how metabolic coupling affects the average correlation between the slow oscillatory components driven by the Poincaré oscillator. Furthermore, Fig. 3(c) illustrates the functional connectivity patterns for the fast oscillatory component generated without (left panel) and with (right panel) metabolic coupling. The increased correlations due to metabolic interactions (highlighted in Fig. 3(a)) lead to more densely connected functional networks in the fast component. In the absence of metabolic coupling, the slow component functional network shown in Fig. 3(d) is sparse (left panel), as there is no coordination, and only cells with very similar intrinsic frequencies are connected. However, the inclusion of metabolic coupling (Fig. 3(d), right panel) facilitates synchronization, resulting in a denser functional network. It should be noted that the same pair-wise thresholds were used for both networks representing the fast (Fig. 3(c)) and slow (Fig. 3(d)) components, respectively.

It is evident that the structures of the two networks differ, with the fast component-derived network bearing more similarities to the structural network (refer to Fig. 1(b)) compared to the slow component-derived network. This finding aligns with experimental observations in pancreatic tissue slices, where the slow component deviates more from the regular structure compared to the fast component network [21,24]. In Fig. 3(c), which presents the fast component network, cell colors indicate the average weights of their connections in the structural network, reflecting thereby junctional conductances. The yellow color represents the highest values, presumably indicating hub cells. Interestingly, it seems that the regions surrounding these cells exhibit the highest density of functional networks. In contrast, the slow component network shown in Fig. 3(d) highlights that in addition to physical distance, the formation of functional connections is also significantly influenced by the similarity of intrinsic metabolic frequencies, which are indicated by the colors of cells.

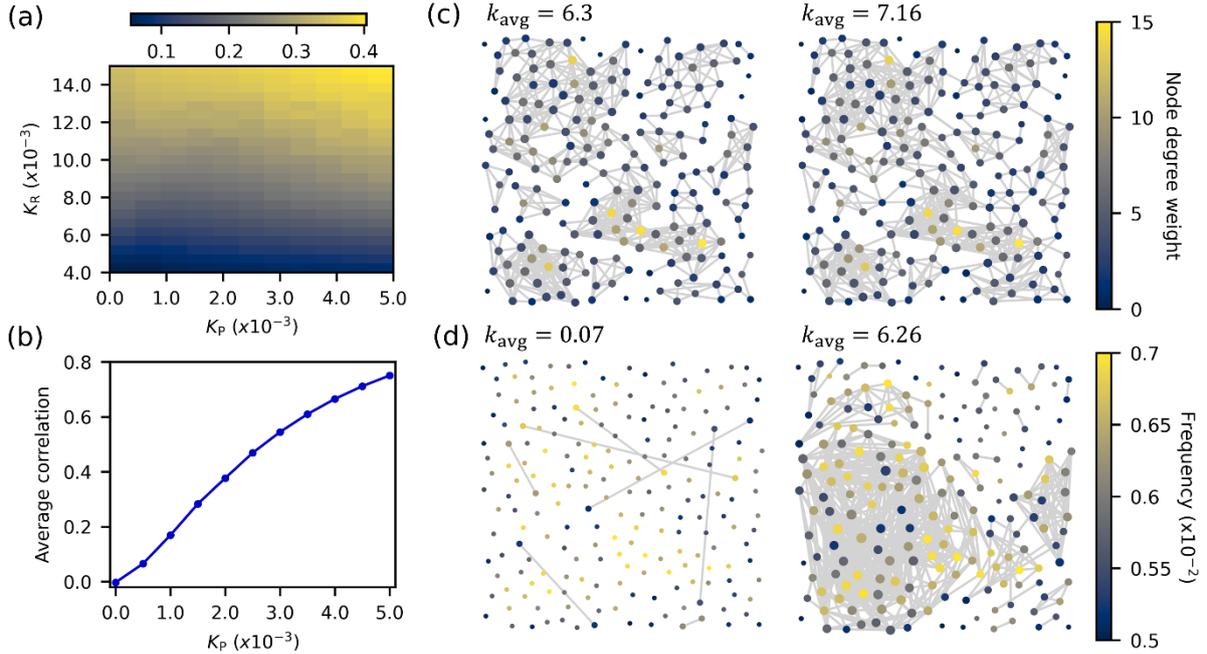

**FIG. 3. The interplay between electrical and metabolic coupling shapes the synchronous behavior and the structure of functional beta cell networks.** (a) Average correlation of the fast-component dynamics as a function of the coupling strength between slow ($K_P$) and fast ($K_R$) oscillatory components. (b) Average correlation of the slow oscillatory activity as a function of slow component coupling strength ($K_P$). (c) Functional networks derived from the fast-component cellular signals without (left, $K_P = 0$) and with metabolic coupling (right, $K_p = 0.004$). For the computation of both networks the same threshold $R_{th}$ was used, and electrical coupling strength was $K_R=0.01$. (d) Functional networks derived from the slow metabolic beta cell activity without (left) and with metabolic coupling (right). In panels (c) and (d) colored dots represent physical locations of cells and grey lines represent functional connections between them, determined from correlations in cellular dynamics. Sizes of dots correspond to the node degrees and colors indicate either the total weighted degree in the structural network (c) or the intrinsic slow-component frequencies (d). In panels (c) and (d) connectivity thresholds ($R_{th}$) were computed based on the left and right panels, respectively, with a target average network node degree $k_{avg}$=6.0 with a 5% tolerance. The same thresholds were then used to design the functional networks on the right (panel (c)) or on the left (panel (d)). On the top of each functional network the average network degree ($k_{avg}$) is indicated.

To evaluate the similarity between network types (structural, fast component-based, and slow component-based), we used coupling parameters $K_R$=0.01 and $K_P$=0.004 in our computations, as they produced behavior like the one observed in experimental measurements of beta cell activity [24,71]. In Fig. 4(a), we quantified the inter-layer similarity coefficient for all network type pairs (see Mathematical model and analysis). Results confirm our previous visual assessment that the Rulkov-driven, i.e., fast component-based, network bears more similarities with the structural intercellular network than the Poincaré-driven, i.e., slow component-based, network. In Fig. 4(b), we compared pairwise the node degree relationship between different networks. In general, there is a positive correlation in degrees between all network pairs, but the relationship is weakest between the structural and the slow degree distributions. The structural network is the most homogeneous, with most cells having 4-6 connections with their neighbors, consistent with what was measured in experiments [68] and with the data shown in Fig. 1(c). Both functional networks are more heterogeneous than the structural network, with node degrees distributed over a broader interval in the slow-component network. Notably, very similar functional connectivity patterns, whose heterogeneity exceeds that of the structural network, were also found in previous experimental and modelling studies [24,42,43,58].

Next, we analyzed how the average correlation between cell pairs is affected by the physical distance between them. The results are presented in Figs. 4(d)-(e). For $K_P>0$, we observe a monotonical decrease of the average correlation between cell pairs as the intercellular distance increases, for both dynamical components, which agrees with experimental measurements [21]. While it is expected that the synchronous behavior of slow oscillations is highly enhanced with increasing values of metabolic coupling (Fig. 4(d)), it is interesting to note that increasing $K_P$ also leads to more coordinated behavior at larger distances in the domain of fast oscillations (Fig. 4(e)). To better understand the principles of the slow functional network, we computed the relative differences in intrinsic metabolic frequencies between connected cell pairs. The results presented in Fig. 4(f) indicate that without metabolic coupling, only cells with very similar frequencies are functionally connected. However, as the coupling strength of the metabolic component increases, the differences in intrinsic frequencies between connected cells also increase. This suggests that the functional connectivity of the slow component is influenced by the interplay of intrinsic metabolic properties and the underlying structural connectivity, indicating thereby why the slow component functional network differs quite profoundly from the structural network. In contrast, the fast oscillatory activity-derived network bears more similarity to the structural connectivity pattern, as it reflects the propagation of intercellular excitation waves that are mediated by structural connections. However, in the regions around the cells with the strongest weights in the structural network, the density of functional connections is very high and seems to exceed the average density of the underlying structural network (Fig. 3(c)). Apparently, the strongest connected cells act as hubs in the functional network, which establishes well-synchronized and interconnected regions around them. In addition to the positive correlation between the node degrees in the structural and functional networks inferred in Fig. 4(b), the relationship between the clustering coefficient and the weighted node degree presented in Fig. 4(g) confirms that the most strongly connected cells that operate as hubs contribute to regionally highly interconnected cellular ensembles.

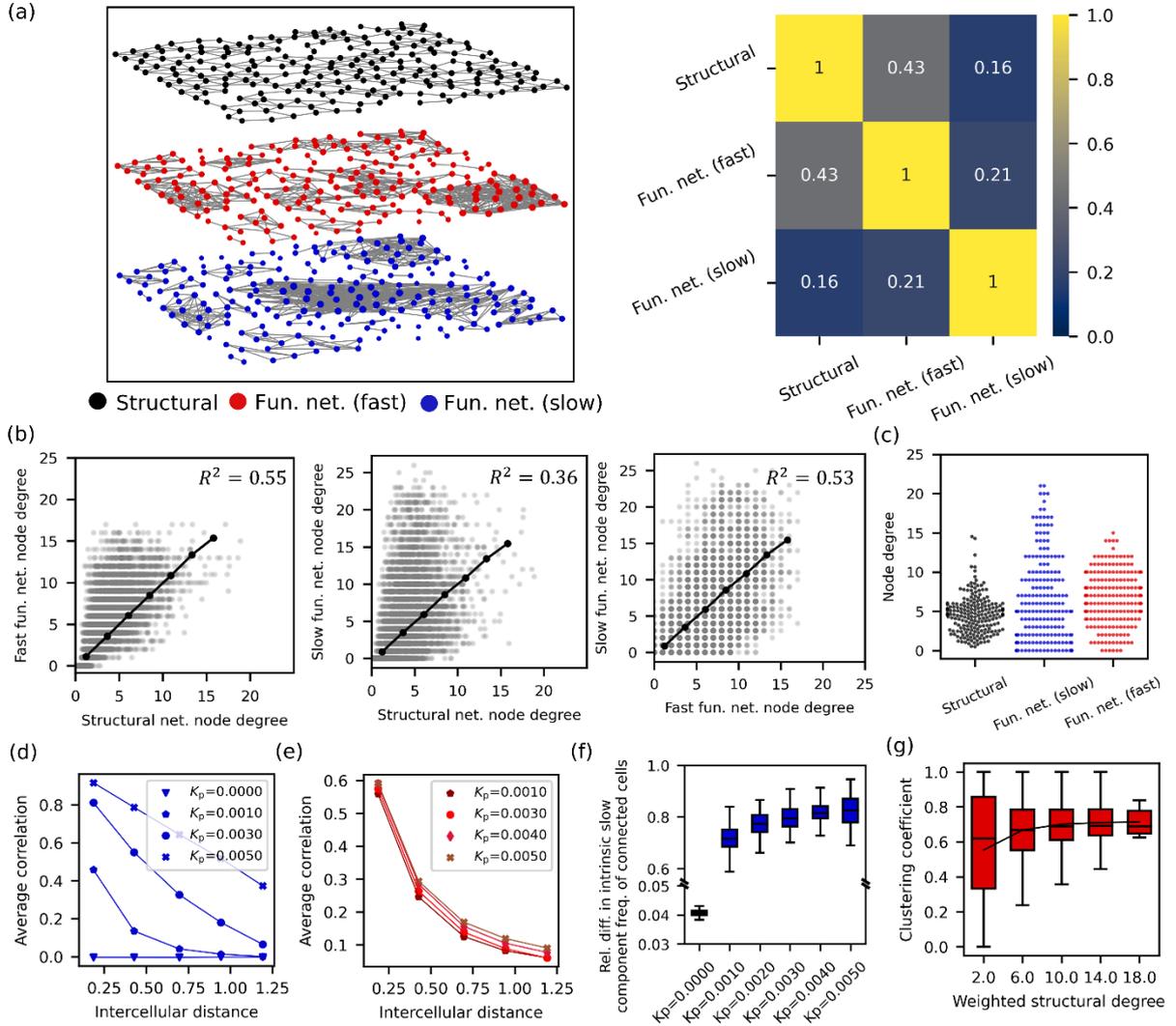

**FIG. 4. Quantifying intercellular synchronicity and networks of slow and fast oscillatory beta cell activity.** (**a**) Multiplex network representation of the structural network (black), and the fast- (red) and slow-component (blue) functional networks (left), and the corresponding interlayer similarity matrix (right). Values in the matrix represent the mean Jaccard similarity between all network-types (see Methods section for further details). (**b**) The pairwise relationships between degrees in different network layers. The grey dots denote values from individual cells and the black dotted line indicates the average trend. Note that the goodness-of-fit $R^2$ is given in individual panels. (**c**) Swarm plots presenting the node degree distribution for the three types of networks. The average correlation as a function of the intercellular distance for different values of metabolic coupling strengths $K_p$ for the slow (**d**) and fast oscillatory component (**e**). (**f**) The relative differences in intrinsic slow-component frequencies between functionally connected cells for different values of $K_P$. (**g**) The clustering coefficient of cells in the fast-activity network as a function of weighted node degree in the structural network. In panels (**f**) and (**g**) the boxes represent the 25th and 75th percentiles, whiskers correspond to the 10th and 90th percentiles, and the black lines indicate the median values. The black line in panel **(g)** corresponds to the average value of the clustering coefficient. In all calculations the fast-component coupling was set to $K_\text{R} = 0.01$. Note that the degree distributions in Fig. 4(c) correspond to the networks in Fig. 4(a), whereas all other calculations presented in Fig. 4 are based on 100 independent simulations on different structural networks. In all functional networks the average node degree was set to 6.

In previous studies, a positive correlation was found between the number of functional connections per beta cell in the fast-dynamics-based network and their activity [16,46]. To investigate how this relationship manifests itself in our multicellular model, we analyzed both the structural and slow component networks. Our simulations show a positive correlation between the node degrees and oscillation frequencies for both networks (Fig. 5), with the strongest correlation observed in the structural network and the weakest in the slow component network. While for the metabolic oscillations a positive correlation between the node degree and the frequency of slow oscillations was expected because it is intrinsically embedded in the model, the relation between the node degree and the frequency of fast oscillations appears to be an emerging phenomenon within the intercellular network. Specifically, coupling facilitates the propagation of intercellular waves, leading to higher cellular activity, especially when excitability is low, i.e., around the nadirs of the slow frequency component. This supports the hypothesis that hub cells play a more significant role in intercellular events than peripheral cells, as coupling is often insufficient to ensure their activation [14]. In our simulations, the frequency differences between the least and most connected cells in the fast-component functional network were around 15-20% (Fig. 5(b)), which is akin to experimental results [14,46]. The relationship between node degrees and the frequencies of slow metabolic oscillations was less pronounced compared to the fast activity in either type of network. However, these relationships, as well as the relations between the cellular activities and the positions in the structural networks, have not yet been evaluated experimentally.

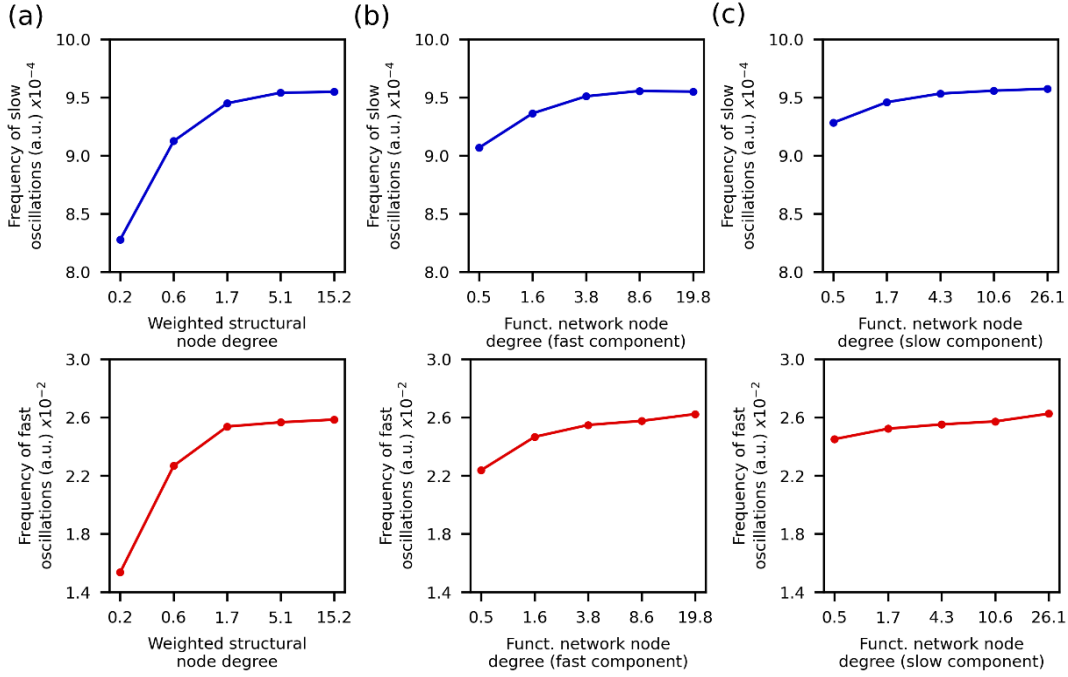

**FIG. 5. Relationship between the oscillation frequencies and the node degrees in different networks.** Frequency of slow-component oscillations (blue lines, upper row) and fast-component oscillations (red lines, lower row) as a function of the weighted node degree in the structural network **(a)**, fast component-derived functional network node degree **(b)** and slow component-derived functional network node degree **(c)**. Individual dots represent the average values for 100 independent simulation runs.

## IV. DISCUSSION

The collective behavior of pancreatic beta cells is characterized by intricate patterns of activity and represents a challenge to study for both experimentalists and computational scientists. In the last decade, advances in high-speed multicellular imaging techniques and complexity science have led to innovative and interdisciplinary approaches to better understand how coherent activity emerges in heterogeneous islet cell populations with the goal to gain further insights into the mechanisms that regulate insulin secretion. In this vein, tools from the armamentarium of the complex network theory applied to multicellular imaging data have become particularly popular and proved themselves as fruitful means to quantitatively describe the intercellular interactions [14,24,42,43,72], identify specific subpopulations [37–41], and to assess how signals are transmitted across the islets [46,73]. However, to date, much of the research in this area has focused on analyzing experimental data, which can offer only limited insights due to the inability to measure all relevant parameters and variables or to fully grasp their interrelationships. To overcome these limitations, computational models represent a viable route to acquire a more mechanistic understanding of how beta cell heterogeneity and their activity profiles shape the functional connectivity patterns, which can complement the knowledge obtained from experimental studies. Until now, there has been limited research that combines computational modelling of beta cell ensembles with functional network analysis [44,45,57,62,74]. In this work, we therefore aimed to contribute to the field by designing a multicellular phenomenological model that incorporates the main features of beta cell dynamics, which we have used to further pursue the abovementioned issues.

The oscillatory activity of beta cells manifests itself on multiple timescales, a feature, that is commonly overlooked by functional network analysis of experimental data. For this reason, we explicitly distinguished in our analyses between the electrically-driven fast and the metabolically-driven slow components of the oscillatory activity. Our findings reveal that the structure of intercellular functional connections differs profoundly between the fast and slow components. Specifically, the fast-component network was found to be less heterogeneous, with fewer long-range connections, and exhibited more similarities with the gap-junctional structural network than the slow-component network. It is worth noting that similar differences between the fast and slow derived functional beta cell networks were observed also in recent experimental studies [21,24]. Furthermore, the structure of the slow-component network was found to be significantly influenced by the intrinsic frequencies that reflect the cells' metabolic activities. This notion is able to clarify the recent and at the first glance arguable findings by Briggs et al. [75], who reported that the functional beta cells networks bear little similarity with the gap-junctional structural network. However, Briggs et al. studied isolated islets that predominantly exhibited slow oscillations and based their analyses on them. Within the framework of our present study, in such a scenario, the functional networks can be expected to be strongly influenced by metabolic dynamics rather than structural coupling. Conversely, our computations indicated that the extracted patterns from the fast activity exhibit a much greater similarity to the structural network. This finding is meaningful because the fast activity is synchronized across islets in the form of intercellular waves that are guided by gap-junctions, as also observed in our simulations (see Supplementary videos S1-S3). However, due to cell-to-cell variability and differences in coupling, the multicellular activity is complex, which leads to some discrepancies in the structures of both networks. Specifically, the functional connectivity network exhibits a higher degree of heterogeneity than one would expect from a syncytium created by gap-junctions and is characterized by the presence of highly connected cells, i.e., hub cells. The emergence of these cells in the functional network is associated with higher levels of gap-junctional conductances, which substantiates previous indirect notion that they have a higher-than-average junctional conductance because they transmit intercellular signals to their neighbor with the shortest delays [46]. Moreover, the presence of hubs can also be associated with the fact that the cells with the highest weights in the structural network form well-synchronized regions around them with a high concentration of functional connections, as also noted in functional connectivity patterns extracted from experimental data [24]. Consequently, the number of functional connections of these cells to other cells can substantially exceed their node degrees in the structural network. In sum, we wish to point out that fast activity-based networks are not a direct reflection of structural networks but can be used as a robust quantitative tool to assess their multicellular dynamics and heterogeneity. Slow activity-based networks importantly differ from fast activity-based networks and future experimental and modelling studies should always explicitly specify which dynamical component their results are based on.

Although the mechanism of synchronizing fast electrical activity across islets via intercellular excitation waves is relatively well understood [32,36,51,76], it remains unclear how slow metabolic-driven oscillatory activity is coordinated among islet cells. Recent experimental data suggest that organized dynamical patterns of the slow component also exist [21,38,39,77]. Considering these findings, we examined in our study how the coupling in the slow component affected the collective activity and the resulting functional connectivity network structures. Our results revealed that the inclusion of metabolic coupling does not only align the slow activity but also contributes to more synchronous dynamics of the fast oscillatory component. The inclusion of interactions in the slow component also enhanced the spatial range of correlated activity and led to denser fast-component functional networks. Furthermore, recent experimental findings indicate that there is not only a notable degree of synchronization of the slow metabolic oscillations, but also that it is a decreasing function of the intercellular distance [21]. We observe the same in our simulations when $K_P>0$.

Importantly, our present modelling study reflects the previously observed experimental behavior of slow oscillations in that they are not completely in phase. Since the fast oscillations are dependent on the slow oscillations and demonstrate a higher frequency and better intercellular synchronization during the peaks of slow oscillations compared with nadirs, stronger metabolic coupling expectedly better aligns the periods of higher and lower activity also in the fast domain. In other words, since our functional networks are based on statistical similarity between fast oscillations in different cells, stronger metabolic coupling increases the similarity between fast oscillations and improves the connectivity of fast component-based networks. More importantly, from a physiological point of view, better synchronicity of the slow component due to better diffusion of glycolytic intermediates, such as glucose-6-phosphate, or a stronger electrical coupling which is able to synchronize the slow component due to feedback from the fast component, or a combination of both, can probably improve the pulsatility of the insulin release at the islet level by better aligning the periods of highest and lowest fast activity in different cells. This pulsatility is in turn crucial for normal beta cell function and responsiveness of target tissues to insulin [9,78]. The fact that blunted pulsatility of insulin release from the islets and changes in insulin oscillations in plasma *in vivo* are an early marker of insulin resistance and diabetes mellitus in both animal models [26,79,80] and human patients [81–83] underlines the importance of understanding the mechanisms underlying the slow oscillations and their relationship with the fast component.

The fast dynamic component has received much more attention in the past and is much better understood in terms of its modulation by secretagogues [3,15,16,19,84], intercellular synchronization [46,51], and network properties [16,38,42,43,46]. Based on the existing evidence and our current findings showing that the modulation by secretagogues [21,24,85], intercellular synchronization [35,76], and network properties [21,24,39] might importantly differ between the slow and the fast component and given the evidence that they are interdependent, the key message of our current study is that future experimental and modelling studies should pay special attention to what dynamic component they are analyzing. Clearly, the findings from the present study need to be validated in the future by experimental approaches, but they also yet again show the importance of modelling studies for understanding the normal and pathological physiology of beta cells and other biological systems with similar complex oscillatory behavior.

**Acknowledgments**

The authors acknowledge the support from the Slovenian Research Agency (research core funding nos. P3-0396 and I0-0029 and research projects nos. J3-3077, N3-0133, and J1-2457).

**Supplementary videos**

**Supplementary video S1.** Raster plot of binarized cellular activity (upper panel) and the corresponding animated activity in the network of coupled cells (lower panel). Colored dots in the lower panel represent physical locations of cells and colors indicate states of cellular activity (red – active, gray – not active). Simulation for coupling parameters: slow-component (Poincaré) coupling $K_P$=0.000, fast-component (Rulkov) coupling $K_R$=0.004.

**Supplementary video S2.** Raster plot of binarized cellular activity (upper panel) and the corresponding animated activity in the network of coupled cells (lower panel). Colored dots in the lower panel represent physical locations of cells and colors indicate states of cellular activity (red – active, gray – not active). Simulation for coupling parameters: slow-component (Poincaré) coupling $K_P$=0.000, fast-component (Rulkov) coupling $K_R$=0.010.

**Supplementary video S3.** Raster plot of binarized cellular activity (upper panel) and the corresponding animated activity in the network of coupled cells (lower panel). Colored dots in the lower panel represent physical locations of cells and colors indicate states of cellular activity (red – active, gray – not active). Simulation for coupling parameters: slow-component (Poincaré) coupling $K_P$=0.004, fast-component (Rulkov) coupling $K_R$=0.010.